# Three-dimensionality of field-induced magnetism in a high-temperature superconductor

B. Lake[1,2], K. Lefmann[3], N. B. Christensen[3], G. Aeppli[4], D.F. McMorrow[4,5], H. M. Ronnow[6], P. Vorderwisch[7], P. Smeibidl[7], N. Mangkorntong[8], T. Sasagawa[8], M. Nohara[8], H. Takagi[9].

[1] Clarendon Laboratory, University of Oxford, Parks Road, Oxford OX1 3PU, U.K.
[2] Ames Laboratory and Department of Physics and Astronomy, Iowa State University, Ames, Iowa 50011, U.S.A.
[3] Materials Research Department, Risø National Laboratory, DK-4000 Roskilde, Denmark.
[4] London Centre for Nanotechnology and Department of Physics and Astronomy, University College London, London WC1E 6BT, U.K.
[5] ISIS Facility, Rutherford Appleton Laboratory, Chilton, Didcot, Oxfordshire OX11 0QX, U.K.
[6] Laboratory for Neutron Scattering, ETH-Zürich and Paul Scherrer Institut, 5232 Villigen, Switzerland
[7] BENSC, Hahn-Meitner Institut, Glienicker Strasse 100, 14109 Berlin, Germany.
[8] Department of Advanced Materials Science, University of Tokyo and CREST-JST, Kasiwa 277-8561, Japan.
[9] RIKEN (The Institute of Physical and Chemical Research), Wako 351-0198, Japan.

**Many physical properties of high-temperature (high-$T_c$) superconductors are two-dimensional phenomena derived from their square planar $CuO_2$ building blocks. This is especially true of the magnetism from the copper ions. As mobile charge carriers enter the $CuO_2$ layers, the antiferromagnetism of the parent insulators, where each copper spin is antiparallel to its nearest neighbours[1], evolves into a fluctuating state where the spins show tendencies towards magnetic order of a longer periodicity. For certain charge carrier densities, quantum fluctuations are sufficiently suppressed to yield static long-period order[2,3,4,5,6], and external magnetic fields also induce such order[7,8,9,10,11,12]. Here we show that in contrast to the chemically-controlled order in superconducting samples, the field-induced order in these same samples is actually three-dimensional, implying significant magnetic linkage between the $CuO_2$ planes. The results are important because they show that there are three-dimensional magnetic couplings which survive into the superconducting state, and coexist with the crucial inter-layer couplings responsible for three-dimensional superconductivity. Both types of coupling will straighten the vortex lines, implying that we have finally established a direct link between technical superconductivity, which requires zero electrical resistance in an applied magnetic field and depends on vortex dynamics, and the underlying antiferromagnetism of the cuprates.**

$La_2CuO_4$ is the parent compound of the original high temperature superconductor $La_{2-x}Ba_xCuO_4$ discovered by Bednorz and Muller[13] nearly two decades ago. Here we examine the related cuprate $La_{2-x}Sr_xCuO_4$ (LSCO) with $x$=0.10 and a superconducting transition temperature of $T_c$=29 K. Previous measurements indicate weak, long-period magnetic order derived from defects which differ from sample to sample, and stronger field-induced order with all of the hallmarks of an intrinsic effect[8], including sample-independence and a sharp onset temperature indistinguishable from $T_c$. While these measurements have stimulated theory[14,15,16], they could only probe magnetism within the $CuO_2$ planes, and thus were insensitive to inter-planar interactions. The latter are ultimately responsible for technologically useful three-dimensional superconductivity and for key features of the competition between magnetism and superconductivity[16]. Apart from the inter-planar correlations, it is important to establish whether it is the coupling of the field to the spins or to the motion of the electrons which is responsible for the field-induced order. Clarification would permit rigorous scrutiny of the idea that the vortices in the superconductors nucleate the field-induced order, a phenomenon rather different from the Zeeman-induced pair-breaking likely to be responsible for the high-field superconductor-insulator transition characteristic of the underdoped cuprates[17,18].

The standard technique for measurement of antiferromagnetic order is neutron diffraction, the magnetic analogue of structural x-ray diffraction. In all previous experiments to characterize the field-induced order in high-temperature superconductors, the field was vertical and approximately perpendicular to the $CuO_2$ planes. The large masses of the magnet and instrumental components prevented them from being rotated about a horizontal axis precluding both the determination of the dependence of the field effect on the angle between the field and the $CuO_2$ planes, and an examination of inter-planar magnetic correlations. Here we pioneer the use of a different neutron scattering geometry and magnet. The field is now horizontal, and the vertical $CuO_2$ planes can be rotated in the magnet around a vertical axis, allowing access to arbitrary angles between the field and $CuO_2$ planes. Furthermore, out-of-plane magnetic correlations can be explored via rotation of instrument components such as the detector arm within the horizontal plane. What makes this experiment much more difficult than previous measurements is the restricted access of the neutron beam to the sample in the more complex magnet (see Fig. 1a and Fig. 1b), the lower magnetic signal associated with the smaller maximum field of the magnet and the imperfect (geometry-related) match between the signal shape (in reciprocal space) and the instrument resolution function.

We used two different LSCO x=0.10 crystals prepared by the same methods as those examined in our previous experiments[8]. Comparison of results for the two crystals from different growth runs allows us to check whether the observed behaviour is intrinsic or extrinsic. The long-period magnetic order in superconducting derivatives of $La_2CuO_4$ manifests itself as a quartet of peaks in reciprocal space with wavevectors $Q_\delta=(1\pm\delta/2,\pm\delta/2,l)$ ($\delta=0.24$)[8]; using the orthorhombic Bmab notation where the **a** and **b** axes span the $CuO_2$ planes and the longer **c** axis is perpendicular to these planes.

We first investigate the dependence of the field-induced signal on field direction. Figure 2a shows a background corrected in-plane scan of sample A through the magnetic peak position of (1.12,0.12,4.0). Measurements were restricted to an inter-layer wavevector of $l$=4.0 due to the limited region of reciprocal space available in the horizontal magnet (Fig. 1b) and the long counting times required (~20 minutes per point). This scan is equivalent by translation of the [0,0,4] reciprocal lattice vector to scans through (1.12,0.12,0) which have been made by several different groups[2,3,4,5,6] on a range of underdoped $La_2CuO_4$-based crystals. In agreement with these findings and our previous results[8], magnetic signal appears in the superconducting state in zero field and is strongly enhanced by external magnetic fields (e.g. $H$=6 T//**c**) perpendicular to the $CuO_2$ planes. However, if the same field has a large component within the $CuO_2$ planes, as for example when the 6 T field makes an angle of 11.5 degrees with the planes, there is much smaller field-induced signal. For this field direction, the in-plane component of field was $H_a$=5.9 T while the component parallel to the inter-planar **c** axis was $H_c$=1.2 T. Figure 2b shows that the reduction on going from **H**=6 T//**c** to the tilted field of the same magnitude is exactly the same as that found in our previous measurements[8] to arise from a reduction of field applied parallel to **c** from 6 T to 1.2 T. The out-of-plane component therefore accounts for the entire field-induced scattering observed for the tilted field, implying that the signal induced by the component of field within the superconducting planes is negligible. One could imagine that a significant contribution to the tilted field effect comes from spin rotation within the planes (to which the spins are confined[5,19]) induced by the 5.9 T in-plane field component, leading to a reduction in the signal since only spin components perpendicular to wavevector transfer **Q** are observed in neutron scattering. However, this is ruled out by the combined facts that **Q** has a large component along **c** and that 5.9 T is below the 7.5-8 T spin-flop fields found for similarly doped LSCO via nuclear magnetic resonance (NMR)[20] and ultrasonic[21] measurements.

The large difference between the magnetism for **H** perpendicular and almost parallel to the inter-layer direction reflects the anisotropies observed in certain other properties of LSCO. Magnetoresistance measurements give the ratio of out-of-plane to in-plane normal state resistivity as $\rho_c/\rho_{ab}$~2000[22], the superconducting coherence length within the $CuO_2$ plane is $\xi_{ab}$~30 Å compared to $\xi_c$~1-2 Å along the **c** axis[23], and ratio of the London penetration depths is $\lambda_c/\lambda_{ab}$~20[24,25]. On the other hand the spin susceptibility as determined e.g. by NMR is isotropic[26]. Our results therefore show that the field-induced signal is due to coupling of the field to the motion of the electrons rather than their spin. Thus, we have very direct evidence that the field-induced order comes from currents circulating within the planes – i.e. vortices – rather than a Zeeman effect. A corresponding anisotropy has been observed in the bilayer cuprate superconductor $YBa_2Cu_3O_{6.6}$ where a field suppresses the magnetic $(\pi,\pi)$ resonance much more strongly if it points perpendicular to the $CuO_2$ planes than parallel to them[27].

We now turn to the inter-planar correlations with the field perpendicular to the $CuO_2$ layers. The top panels in Fig. 3 show raw data for an in-plane scan through the magnetic peak position of (1.12,0.12,4.0) at a fixed out-of-plane wavevector $l$=4.0 for both samples A and B. The bottom two panels show a similar scan collected away from the reciprocal lattice point at $l$=3.6. For sample A, the zero-field signal at this wavevector is greater than at $l$=4.0; but, application of a magnetic field produces negligible change in the magnetic intensity. For sample B, the zero-field signal at $l$=3.6 is similar to that at $l$=4.0, and a 6 T field again has no impact. Analysis of the lineshape yields a resolution-limited in-plane correlation length of $\zeta_{in\text{-}plane}$>200 Å for both samples at all temperatures and fields measured, confirming that there are long-range magnetic correlations within the $CuO_2$ layers.

To establish the magnetic structure factor perpendicular to the planes, we made similar scans to those shown in Fig. 3 for all other values of $l$ allowed by the restricted experimental geometry (Fig. 1a and Fig. 1b). The full data sets for both samples are given in Fig. 4. There are several startling features. First, for sample A, the signal is fairly constant far from $l$=4.0, but at $l$=4.0 it passes through a minimum. We do not however observe such an effect for sample B, for which the signal seems $l$-independent. These contrasting results again suggest that the zero-field magnetism is extrinsic and due to sample-dependent defects (see supplementary material section 1) which in some cases yield scattering with a minimum at (0,0,4), as in sample A (see supplementary material section 2)[28]. The difference at $T$=2 K between the peak amplitudes at $H=6$ T and zero field, is plotted in Fig. 4d to give the field-induced amplitude in the superconducting state. Even though the zero-field data for the two samples appear very different, a magnetic field of $H=6$ T applied perpendicular to the $CuO_2$ planes significantly enhances the scattering around $l=4.0$ so as to produce a relatively sharp peak in the difference scans for both samples. The applied field therefore enhances the inter-planar correlations so that while the zero-field signal is two-dimensional, the field-induced signal is much more three-dimensional. The difference scans can be used to estimate these correlation lengths ($L$) as 15.8±2.2 Å and 23.9±2.8 Å for the field-induced order in samples A and B respectively. When the errors in the measurements are taken into account, the two lengths are consistent with each other and correspond to correlated regions extending over approximately 6 $CuO_2$ planes (given by $2L/d$ where $d=6.6$ Å is the inter-planar separation equal to c/2 since there are two planes per unit cell).

Due to the very small superconducting coherence length along the **c** axis of $\xi_c \sim$1-2 Å[23] for LSCO, which is less than the inter-planar distance $d$, vortices in the cuprate superconductors should be regarded as two-dimensional disks or 'pancakes'. Couplings regulating the stacking of such pancakes are therefore important for the establishment of phase coherent superconductivity in applied magnetic fields[29,30], and also in determining three-dimensional flux flow properties, which depend strongly on the alignment from layer to layer of the pancake vortices to form three-dimensional vortex lines. Our experiments reveal two important facts. The first is that antiferromagnetism arises from the coupling of the external field to orbital currents in the $CuO_2$ layers. The second is that the field-induced antiferromagnetism displays significant three-dimensionality. Together they imply that the magnetic fields characteristic of technical applications of superconductivity can induce antiferromagnetism nucleated by vortices, and that bending energies of the vortex lines will contain terms derived from these antiferromagnetic couplings in addition to the well-known Josephson couplings between layers. In particular, inter-planar antiferromagnetism will act to straighten vortex lines because this would reduce their energy by producing greater numbers of antiferromagnetically correlated spins when pancakes in neighbouring planes are aligned over each other (see Fig. 1d). The Josephson couplings, as deduced from optical conductivity data[31] on actual LSCO superconductors, are around 1 μeV, while the interlayer antiferromagnetic couplings, currently established only for the parent and lightly doped compounds, are also of order 1 μeV[32]. Because the Josephson energies are comparable to the antiferromagnetic couplings, we expect the interplanar magnetism to have a significant effect on the vortex bending and vice versa.

Just as they provide the first evidence that antiferromagnetism influences technical superconductivity, our experiments also show that technical superconductivity impacts the more esoteric world of the quantum phase transitions which some believe to be crucial to the mechanism for high-$T_c$ phenomenon. Indeed, they confirm that the magnetic field is a singular perturbation, qualitatively different from chemical composition and pressure. For pressure or doping, the standard picture is that magnetic order is maintained until some threshold, frequently referred to as a quantum critical point, is reached. Well beyond the dopings or pressures required for the zero-field quantum critical points, field-induced order can appear for infinitesimal fields. This follows because a field induces vortex lines which in turn nucleate antiferromagnetism in consecutive planes simultaneously, thus immediately making the coupling between planes relevant and inducing strong magnetic order of a type not brought about by point defects[16].

**Acknowledgements**
We thank N. H. Andersen, B. M. Andersen, A. B. Abrahamsen and S. Kivelson for valuable discussions. Experiments at the Berlin Neutron Scattering Centre were made possible by the support of the European Community - Access to Research Infrastructure action of the Improving Human Potential Programme. Work in London was supported by Royal Society Wolfson Merit Awards.

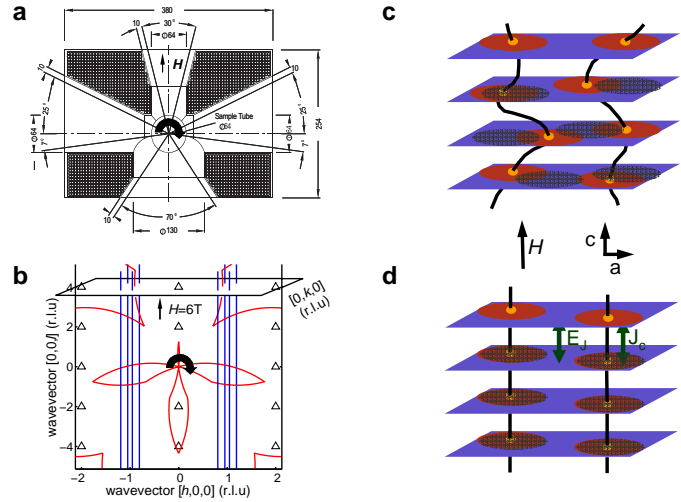

**Figure 1** The experimental configuration and real-space interpretation of the results. (a) shows a cross-section of the horizontal-field magnet (HM1) in the horizontal neutron scattering plane of the V2/FLEX triple axis spectrometer located at the BER-II reactor, Hahn-Meitner Institute (HMI), Berlin, Germany. This diagram was taken from the HMI sample environment manual. The magnet has four windows through which neutrons can pass, separated by four blind spots (hatched regions) which are opaque to neutrons. The LSCO crystal lies at the centre of the magnet in the sample tube and can be rotated within the magnet about the vertical (grey arrow) thus changing the orientation of the crystal axes with respect to the field direction. (b) shows the accessible wavevectors (regions enclosed by the red lines) for magnetic field pointing along the c axis, neutrons with energy of 12.95 meV and a maximum scattering angle of 60 degrees. The crystal is twinned below the tetragonal-to-orthorhombic transition temperature of $T_{t \to o}(x=0.10)=275$ K, but the present measurements do not resolve the twins. Here the reciprocal space for the $(h,0,l)$-domain is illustrated and the black triangles give the positions of the nuclear Bragg peaks. Magnetic scattering (blue lines) is observed at low temperatures as four incommensurate peaks in the $CuO_2$ planes surrounding (1,0,0). (c) and (d) show schematics of the superconducting and magnetic order in real space. The sheets represent the $CuO_2$ planes and the orange circles represent the 'pancake' vortices for magnetic fields applied perpendicular to the $CuO_2$ planes. The red disks represent the magnetic regions nucleated by vortices and the grey shadows of the magnetic patches on the plane below indicate the degree to which they are stacked along the c axis. $E_J$ and $J_c$ are the magnetic and Josephson coupling energies. Entropy and defects favour random stacking of the pancakes, implying that the vortex lines joining pancakes in adjacent planes would not be straight (part (c)) and the magnetic regions nucleated by vorticies or defects would also not be aligned along c. Josephson coupling between the superconducting regions outside the pancakes on neighbouring layers favours straightened vortex lines because the energy would be lowered if the superconducting regions in adjacent planes are aligned. In LSCO, three-dimensional magnetism is associated with the vortices, implying an additional, magnetic energy reduction from stacking the pancakes on top of each other in the c direction and so further straightening the 'vortex lines' (part (d)).

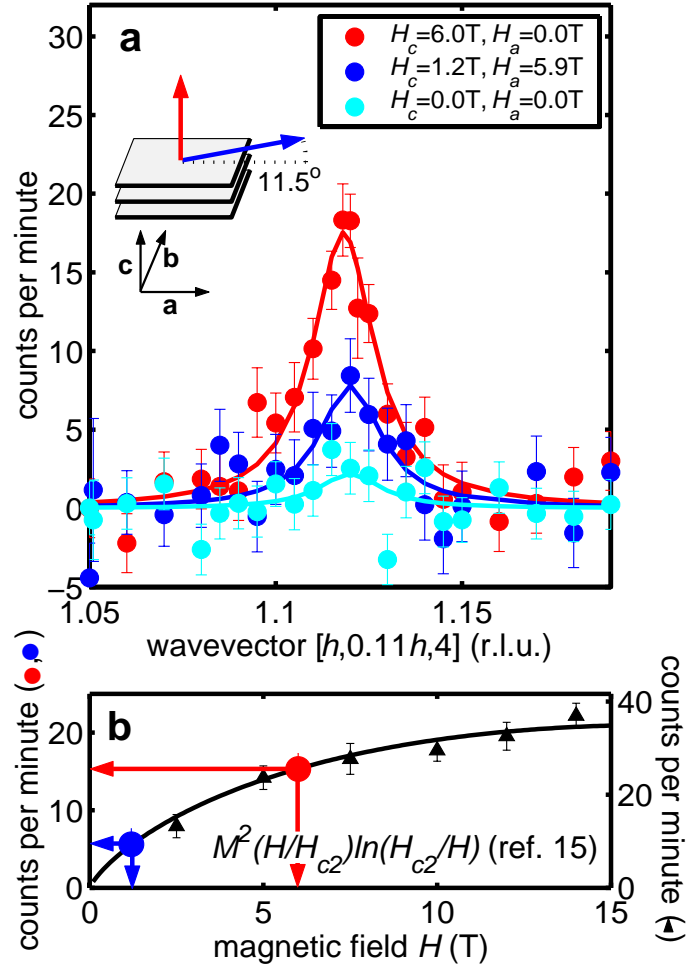

**Figure 2** Field-induced signal in sample A for two different field directions. (a) shows measurements made at (1.12,0.12,4.0). The light blue, red, and dark blue symbols represent the background-subtracted signal observed at 2 K with fields of zero, 6 T pointing perpendicular to the CuO$_2$ planes ($H$=6 T//c), and 6 T applied at an angle 11.5 from the planes (inset), respectively. The lines through the data are fits of a Lorentzian lineshape convolved with the full four-dimensional resolution function. (b) shows field-induced peak amplitudes from the current experiment, using colour coding as in (a), together with previous data, collected for $H$//c but where the CuO$_2$ planes of LSCO formed the horizontal scattering plane of the neutron spectrometer (black filled triangles)[8]. Because of the reduced efficiency of the new scattering geometry, the count rates are lower in the current experiment. The solid black line represents the theoretical expression for the field-induced signal derived by E. Demler *et al*[15] for a field parallel to c.

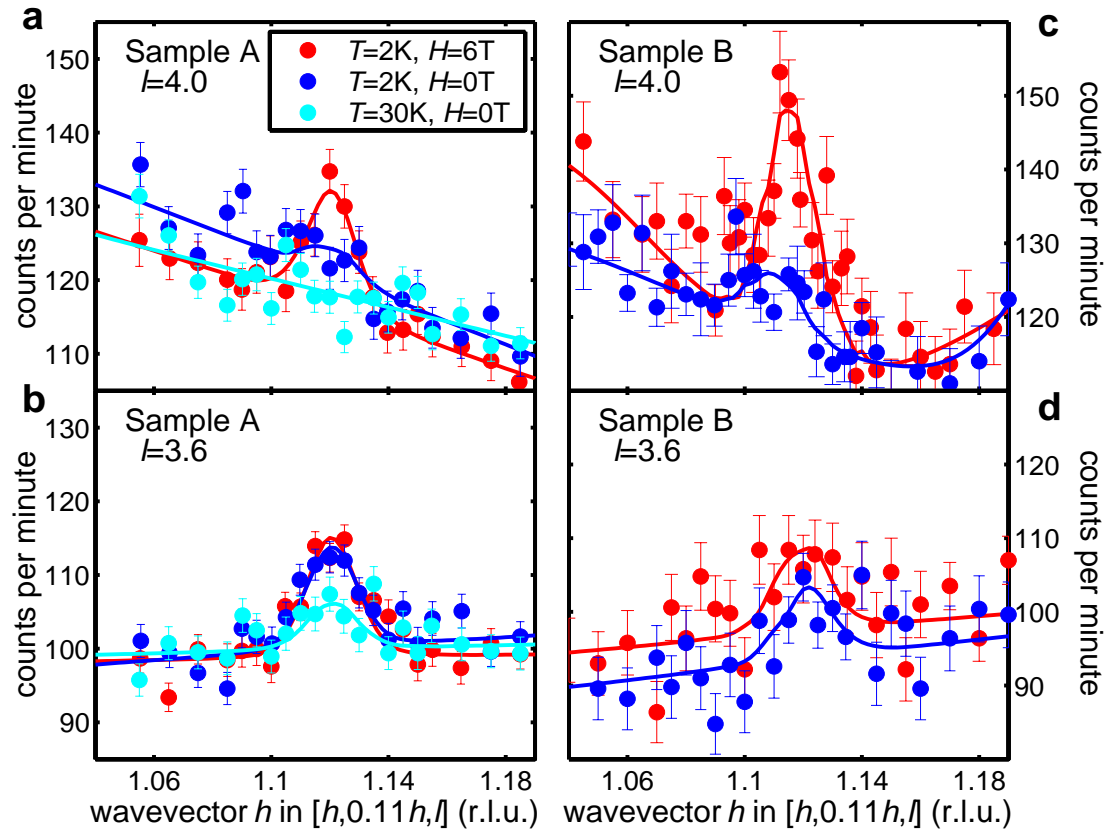

**Figure 3** Magnetic scattering from LSCO x=0.10 at different fields and temperatures for the two different samples. The light blue, dark blue, and red symbols represent uncorrected data collected in the normal state ($T$=30K) in zero field, the superconducting state ($T$=2 K) in zero field, and the superconducting state ($T$=2 K) for a 6 T field applied perpendicular to the $CuO_2$ plane ($H$//c), respectively. (a) and (b) show data collected from sample A, while (c) and (d) show the same scans for sample B. The top two panels ((a) and (c)) show data collected at (1.12,0.12,4.0) while the bottom two panels ((b) and (d)) show data collected at (1.12,0.12,3.6). The lines through the data are fits of a Lorentzian lineshape convolved with the full response function of the instrument. A collimation of guide-60'-60'-open was used to define the divergence of the neutron beam and provide an instrumental resolution of 0.028 $Å^{-1}$. The neutron energies were fixed at 12.95 meV using a pryrolytic graphite (PG) monochromator and analyser, and a tuneable PG filter was used to eliminate second-order scattering.

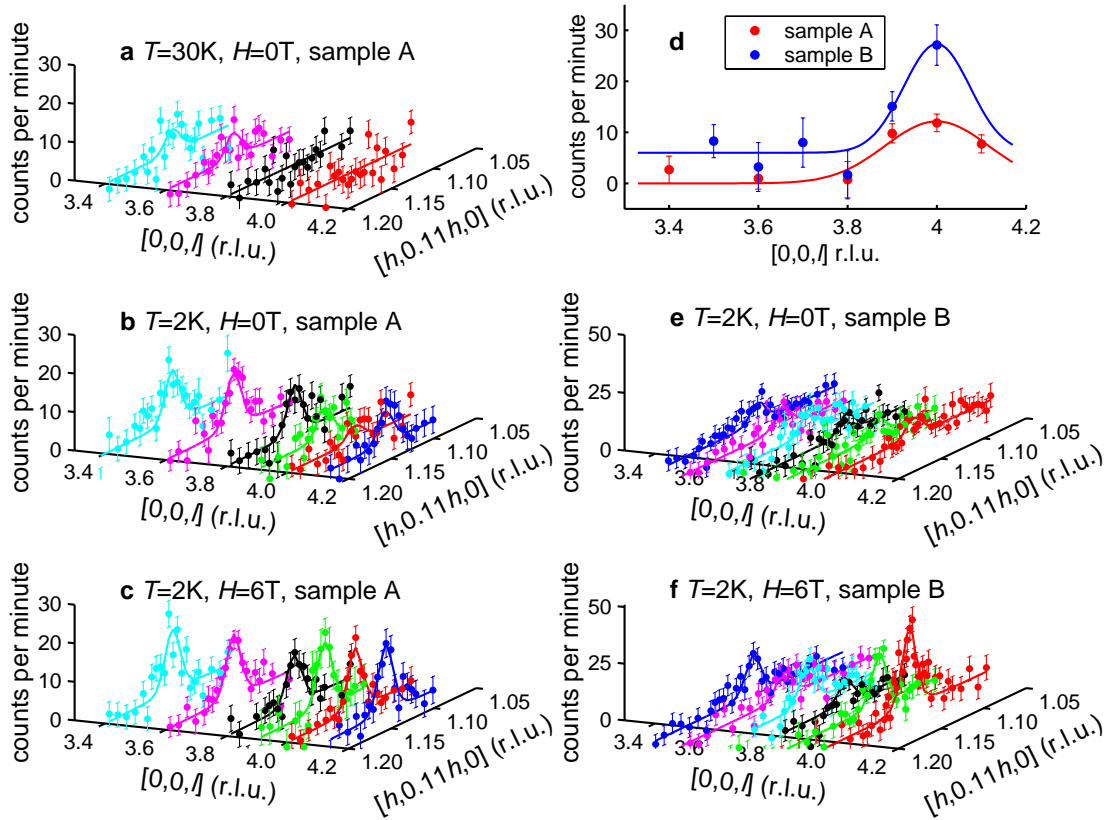

**Figure 4** Data collected for both samples as scans where the in-plane wavevector was varied for various fixed interplanar wavevectors; each scan has been corrected for background and is represented by a different colour. The scan ranges are severely limited by the magnet supports as shown in Fig. 1a and Fig. 1b. The coloured lines through the data are from fits of a Lorentzian profile convolved with the full four-dimensional resolution function and in all cases the peak was found to be resolution-limited within the $CuO_2$ planes. (a), (b) and (c) give the data collected for sample A for $T=30$ K $H=0$ T, $T=2$ K $H=0$ T, and $T=2$ K $H=6$ T respectively; while (e) and (f) give the data for sample B at $T=2$ K $H=0$ T, and $T=2$ K $H=6$ T. (d) shows the amplitude of the field-induced magnetic scattering in the superconducting state (difference between the 6 T and 0 T amplitudes for $T=2$ K) for both sample A (red circles) and sample B (blue circles). The lines through the data are Gaussians superimposed on flat signals.